\documentstyle[preprint,aps,pre,eqsecnum,epsfig]{revtex}
\draft
\newcommand{\h}{\mathcal{H}}
\newcommand{\ds}{\displaystyle}
\def\beq{\begin{equation}}
\def\eeq{\end{equation}}
\def\beqa{\begin{eqnarray}}
\def\eqa{\end{eqnarray}}
\def\r2{$\sqrt{2}$}

\begin{document}

\title{The Anisotropic Ashkin-Teller Model: a Renormalization
Group Study}

\vskip \baselineskip

\author{C.G. Bezerra$^{1}$, A.M. Mariz$^{1}$, J.M. de Ara\'{u}jo$^{2}$ and F.A. da Costa$^{1}$}

\address{
$^1$Departamento de F\'{\i}sica Te\'orica e Experimental \\
Universidade Federal do Rio Grande do Norte \\
Campus Universit\'{a}rio -- Caixa Postal 1641 \\
59072-970 \hspace{8mm} Natal - RN \hspace{8mm} Brazil\\
$^2$Departamento de F\'{\i}sica  \\
Universidade do Estado do Rio Grande do Norte \\
59610-210 \hspace{8mm} Mossor\'{o} - RN \hspace{8mm} Brazil}

\date{\today}
\maketitle

%\newpage
%\vskip \baselineskip

\begin{abstract}
The two-dimensional ferromagnetic anisotropic Ashkin-Teller model is investigated
through a real-space renormalization-group approach. The critical frontier,
separating five distinct phases, recover all the known exacts results for the
square lattice. The correlation length $(\nu_T)$ and crossover $(\phi)$ critical
exponents are also calculated. With the only exception of the four-state Potts
critical point, the entire phase diagram belongs to the Ising universality class.

\vskip \baselineskip
%\vspace{2cm}
%\medskip



\end{abstract}

\newpage

\section{Introduction}

The Ashkin-Teller (AT) model was introduced to study cooperative
phenomena of quaternary alloys \cite{ashtel} on a lattice.
Lately it was shown that it could be described in a hamiltonian form
appropriated for magnetic systems \cite{fan}. In this representation
the AT model can be considered as two superposed Ising models described
by classical spins variables $\sigma$ and $\tau$. In addition, the Ising
systems are coupled by a four-spin interaction term. The isotropic AT
(IAT) model corresponds to the case where the two Ising systems are
identical to each other.

The phase diagrams for the IAT model are relatively well known in both two
and three dimensions. Many exact results were obtained for the IAT on a
square lattice \cite{baxter,nienhuis}. The corresponding phase diagram was
determined by real-space renormalization group analysis \cite{mariz85},
mean-field renormalization-group approach \cite{pla,pmco,bjp99}, Monte-Carlo
renormalization group \cite{chahine} and Monte Carlo simulations
\cite{wiseman,kamieniarz}. These results may be compared to the phase
diagram of the selenium adsorbed on $Ni$ surface, which is a good physical
realization for the two-dimensional IAT \cite{bak,ochab}. The oxigen
ordering in $YBa_{2}Cu_{3}O_{z}$ may also be understood in analogy with the
two-dimensional IAT model \cite{bartI,bartII,wille}. The three-dimensional
IAT model was extensively investigated by a number of techniques
\cite{ditz}, showing an enormous richness of critical behaviour. Mean-field
calculations suggest the existence of many multicritical points, some of
them confirmed by Monte Carlo simulations \cite{ditz}. A Cayley tree
formulation suggests that, for high enough coordination, there may be also an
additional phase in the antiferromagnetic IAT where a symmetry
breaking between the spins $\sigma$ and $\tau$ takes place \cite{dacosta} .

The physical description of the anisotropic AT model (AAT), however, is not
so well known. Using exact duality relations Wu and Lin \cite{wulin}
determined the general topology of the phase diagram for the AAT on a
square lattice. Their results were lately supported by a Migdal-Kadanoff
renormalization group study \cite{domany}. Recently, Benyoussef and
colaborators have investigated the AAT using both mean-field treatment and
Monte Carlo simulations \cite{bek}, as well as finite-size scaling
\cite{bad}. In this paper we use a renormalization-group approach which,
despite its simplicity, gives very accurate results for Ising-like models
on the square lattice \cite{tsallis81,mariz85}, for the resistor network
\cite{tsallis83a} and directed percolation \cite{tsallis83b}. Motivated by
these results we expect that, within its validity limits, the present
approach gives good estimates for the critical properties of the AAT,
mainly for the phase boundaries. In the next section we define the model and
obtain the recursion relations. In section III the results of our analysis
are presented and compared with known results. Finally, in section IV we
summarize our findings.

\section{The model and the renormalization group equations}

The hamiltonian of the Ashkin-Teller model on a squate lattice is given by

\beq
{\h} = -J_{1} \sum_{\langle ij \rangle} \sigma_{i}\sigma_{j}
           -J_{2} \sum_{\langle ij \rangle} \tau_{i}\tau_{j}
           -J_{4} \sum_{\langle ij \rangle}
           \sigma_{i}\tau_{i}\sigma_{j}\tau_{j}
~, \label{ham}
\eeq
where $\sigma_{i}=\pm1$ and $\tau_{i}=\pm1$, $J_{1}$ is the coupling
between the spin variables $\sigma$, $J_{2}$ is the coupling between the
spin variables $\tau$, and $J_{4}$ represents the four-spin interaction
that couples the two Ising systems. The sums $\sum_{\langle ij \rangle}$
runs over all first-neighbors pairs of sites.

The model described by (\ref{ham}) will be studied by a real-space
renormalization-group approach based on the self-dual graph shown in Fig.\
1. It is well-known\cite{mariz85,tarev} that the method is exact for classical spin models
defined on the hierarchical lattice generated iterating the graph of Fig.\ 1, and produces a
very good numerical approximation for the phase diagram of the square lattice. Since our approach does not allow any sublattice structure
such as anti-ferromagnetic ordering we will require that the couplings
satisfy the following ferromagnetic condition

\beq
J_{1}+J_{2} \ge 0, \qquad J_{1}+J_{4} \ge 0, \qquad J_{2}+J_{4} \ge 0 .
\eeq

The renormalization is performed by the decimation of the spin variables
associated with the sites 3 and 4 in Fig. 1. To simplify the calculations
is convenient to introduce the transmissivity vector
${\bf t} = (t_{1},t_{2},t_{3})$ \cite{tarev,alcaraz} through

\beqa
t_{1} & = &
\ds{\frac{1-\exp(-2K_{1}-2K_{2})+\exp(-2K_{1}-2K_{4})-\exp(-2K_{2}-2K_{4})}
{1+\exp(-2K_{1}-2K_{2})+\exp(-2K_{1}-2K_{4})+\exp(-2K_{2}-2K_{4})}} , \\
t_{2} & = &
\ds{\frac{1+\exp(-2K_{1}-2K_{2})-\exp(-2K_{1}-2K_{4})-\exp(-2K_{2}-2K_{4})}
{1+\exp(-2K_{1}-2K_{2})+\exp(-2K_{1}-2K_{4})+\exp(-2K_{2}-2K_{4})}} , \\
t_{3} & = &
\ds{\frac{1-\exp(-2K_{1}-2K_{2})-\exp(-2K_{1}-2K_{4})+\exp(-2K_{2}-2K_{4})}
{1+\exp(-2K_{1}-2K_{2})+\exp(-2K_{1}-2K_{4})+\exp(-2K_{2}-2K_{4})}},
\eqa
where $K_{i} = J_{i}/k_{B}T = \beta J_{i}$ ($i=1,2,4$).

It can be shown that in the renormalization of a series array of two bonds
the vector ${\bf t}$ is simply given by the product of the corresponding
vectors of the bonds \cite{tarev,alcaraz}. In a similar way, in the renormalization of a
parallel array of two bonds the same result holds for the dual vector ${\bf
t}^{D}$, whose components are given by

\beqa
t_{1}^{D} & = & \frac{1+t_{1}-t_{2}-t_{3}}{1+t_{1}+t_{2}+t_{3}} , \\
t_{3}^{D} & = & \frac{1-t_{1}+t_{2}-t_{3}}{1+t_{1}+t_{2}+t_{3}} , \\
t_{1}^{D} & = & \frac{1-t_{1}-t_{2}+t_{3}}{1+t_{1}+t_{2}+t_{3}} .
\eqa

The renormalization-group equations are obtained by requiring that the
partition function remains invariant after the decimation of the
intermediate spins siting on the vertices 3 and 4 represented in Fig. 1(a).
The is achieved by writing

\beq
\exp(-\beta {\cal H}^{\prime}_{1,2}+K_{0}^{\prime}) = \sum_{\{\sigma_{3},
\tau_{3}\}}
\sum_{\{\sigma_{4},\tau_{4}\}} \exp(-\beta {\cal H}_{1,2,3,4})
\eeq

\noindent
where ${\cal H}^{\prime}_{1,2}$ and ${\cal H}_{1,2,3,4}$ are the cluster
hamiltonians associated with Figs. 1(a) and 1(b), respectively, and
$K_{0}^{\prime}$ is an additive constant.

It follows that the recursion relations for the renormalization group thus
defined are given by

\beqa
t^{\prime}_{1} & = &
R(t_1^2+t_1^3+t_1t_2^2t_3^2+t_2^2t_3^2+2t_1t_2t_3^2+2t_1t_2^2t_3) ,
\label{rg1a} \\
t^{\prime}_{2} & = &
R(t_2^2+t_2^3+t_1^2t_2t_3^2+t_1^2t_3^2+2t_1t_2t_3^2+2t_1^2t_2t_3) ,
\label{rg1b} \\
t^{\prime}_{3} & = &
R(t_3^2+t_3^3+t_1^2t_2^2t_3+t_1^2t_2^2+2t_1t_2^2t_3+2t_1^2t_2t_3)
\label{rg1c}
\eqa
where

\beq
R = 2(1+t_1^4+t_2^4+t_3^4+2t_1^3+2t_2^3+2t_3^3+2t_1t_2^2t_3^2+
2t_1^2t_2t_3^2+2t_1^2t_2^2t_3)^{-1} .
\eeq

These equations completely determine the phase diagram for the AAT model,
as well as the thermal critical exponents. Before we proceed with the
analysis of the phase diagram, let us consider another set of variables
defined trough

\beqa
X & = & {\ds \frac{t_1^D + t_3^D}{2}} , \\ Y & = & t_2^D , \\ Z & = & t_1^D
- t_3^D .
\eqa

Using the above introduced variables, the recursion relations given by Eqs.
(\ref{rg1a}-\ref{rg1c}) become

\beq
X^{\prime} = \frac{N_1}{D}, \quad Y^{\prime} = \frac{N_2}{D}, \quad
Z^{\prime} = \frac{N_3}{D}, \label{rge}
\eeq
where

\beqa
N_1 & = & 16X^3X^2+32X^3Y+16X^3+48X^2Y^2+16X^2-4XY^2Z^2-8XYZ^2
\nonumber \\
    & & +12XZ^2- 4Y^2Z^2 + 4Z^2 ,  \label{rg2a}
\\ N_2 &=&
16X^4Y+16X^4+64X^3Y-8X^2YZ^2-8X^2Z^2-16XYZ^2+16Y^3 \nonumber \\
    & & +16Y^2+YZ^4+Z^4 ,  \label{rg2b}
\\
N_3 & = & 48X^2Z-16X^2Y^2Z-32X^2YZ-32XY^2Z+32XZ+4Y^2Z^3 \nonumber \\
&&+8YZ^3+4Z^3 , \label{rg2c}
\\ D &
= & 16X^4Y+16X^4+32X^3Y^2+32X^3-8X^2YZ^2+24X^2Z^2-8XY^2Z^2 \nonumber \\
&& +24XZ^2+8Y^4 +16Y^3+YZ^4+Z^4+8 .
\label{rg2d}
\eqa

The new variable $Z$ gives a measure of the anisotropy (essencially the
difference between $K_1$ and $K_2$), and vanishes in the isotropic limit.
We also note the Eqs. (\ref{rge}-\ref{rg2d}) produce a symmetric flux under a
reflection with respect to the $Z = 0$ plane,
in such a way that the regions $Z>0 ~~ (K_2>K_1)$ and
$Z<0 ~~ (K_2<K_1)$ are isomorphous one to the other. For this reason we
will restrict ourselves to discuss the subspace defined by

\beq
0 \le X \le 1, \quad 0 \le Y \le 1 , \quad Z \ge 0.
\eeq

In the next section we will analyse the fixed points of our recursion
relations, their respective thermal exponents and present the resulting
phase diagram.

\section{The phase diagram}

The renormalization group expressed by Eqs. (\ref{rg1a})-(\ref{rg1c})
reveals the existence of five phases separated by two-dimensional critical
surfaces: {\bf(i)} Paramagnetic $P$ ($<\sigma>=<\tau>=<\sigma \tau>=0$);
{\bf(ii)} Ferromagnetic $F$ ($<\sigma>\neq0,<\tau>\neq0,<\sigma \tau>\neq0$);
{\bf(iii)} Intermediate $I$ ($<\sigma>=<\tau>=0,<\sigma \tau>\neq0$);
{\bf(iv)} Ferro-sigma $F_\sigma$ ($<\sigma>\neq0,<\tau>=<\sigma \tau>=0$); and
{\bf(v)} Ferro-tau $F_\tau$ ($<\tau>\neq0,<\sigma>=<\sigma \tau>=0$). The resulting
phase diagram in the space variables $t_1^D, t_2^D$ and $t_3^D$ is shown in Fig.\ 2,
and reproduces qualitatively the
results presented by \cite{wulin}, \cite{domany} and \cite{bad}. The
critical surfaces represent the boundaries of the domains of attraction of
the trivial fixed points. On these surfaces we find nine (non-trivial)
critical fixed points, which are presented in Table \ref{tab1} along with their
respective critical exponents $\nu_T$ and $\phi$ determined from the
scaling factor $b$ and the relevant eigenvalues \cite{tarev}.
The location of the non-trivial fixed points recovers all
the available exact results for the square lattice. The fixed point $Potts$
corresponds to the four-state Potts model and is completely unstable,
with three relevants eigenvalues and two equal crossover exponents.
All other non-trivial fixed points ($I_1$ through $I_9$) are on the Ising
universality class. Among these points, $I_1$, $I_4$ and $I_7$ have two
relevant eigenvalues. In particular, it is well-known that $I_1$ is not in
the Ising universality class \cite{mariz85}; we expect that the same is
true for $I_4$ and $I_7$. The remaining non-trivial fixed points have a
single relevant eigenvalue, and so they belong to the Ising universality
class. As in the IAT, all phase transitions are second- or higher-order
ones.

\begin{table}
\begin{tabular}{|c|c|c|c|c|} \hline
Fixed point  &  ($t_1^D,t_2^D,t_3^D$)   &   ($X,Y,Z$)  &   $\nu_T$  &  $\phi$ \\ \hline
$Potts$ & (1/3, 1/3, 1/3)  &   (1/3, 1/3, 0) & 0.948  & 2.73 \\ \hline
$I_1$ & (\r2-1, 3-2\r2 , \r2-1) & (\r2-1,3-2\r2,0) & 1.149 & 1.00 \\ \hline
$I_2$ & (0,\r2-1,0) & (0,\r2-1, 0) & 1.149 &  \\ \hline
$I_3$ & (\r2-1,1,\r2-1) & (\r2-1,1,0) & 1.149 &  \\ \hline
$I_4$ & (\r2-1,\r2-1,3-2\r2) & ($\frac{2-\sqrt{2}}{2}$,\r2-1,3\r2-4) & 1.149 & 1.00  \\ \hline
$I_5$ & (\r2-1,0,0) & ($ \frac{\sqrt{2}-1}{2}$,0,\r2-1) & 1.149 &  \\ \hline
$I_6$ & (1,\r2-1,\r2-1) & ($\frac{\sqrt{2}}{2}$,\r2-1,2-\r2) & 1.149  &  \\ \hline
$I_7$ & (3-2\r2,\r2-1,\r2-1) & ($\frac{2-\sqrt{2}}{2}$, \r2-1,4-3\r2) & 1.149 & 1.00 \\ \hline
$I_8$ & (0,0,\r2-1) & ($\frac{\sqrt{2}-1}{2}$,0,1-\r2) & 1.149 & \\ \hline
$I_9$ & (\r2-1,\r2-1,1) & ($\frac{\sqrt{2}}{2}$,\r2-1,\r2-2) & 1.149 & \\ \hline
\end{tabular}
\caption{Non-trivial critical fixed points and critical exponents
$(\nu_T,\phi)$. For the square lattice the exact values are
$\nu_{T}=2/3$ and $\nu_{T}=1$ for the Potts and Ising universality class,
respectively.} \label{tab1}
%\caption{Fixed points and respective critical exponents for the anisotropic Ashkin-Teller model}
\end{table}

Fig. 3 shows the phase diagram in the variables $X$, $Y$ and $Z$.
As mentioned in the preceeding section, it suffices to consider the
region $Z>0$. This symmetry corresponds to a permutation of the $\sigma$ and
$\tau$ spin variables ($F_\sigma$ and $F_\tau$ phases). In this diagram there
is a self-dual plane which is invariant under the renormalization group transformation
and is given by

\beq
2X+Y+Z = 1 .
\eeq

This self-dual plane contains two lines of fixed points. One of these lines
is the intersection of the self-dual and the $Z = 0$ planes, corresponding
to the IAT model and contains the $F-P$ boundary. The other line is
determined by the intersection of this self-dual plane with the
other one given by

\beq
2X + 4Y - Z = 2, \quad Z>0 ,
\eeq

\noindent
and corresponds to the merging of the critical surfaces $P-I$, $I-F$,
$F-F_{\sigma}$ and $F_{\sigma}-P$.

We believe that these results represent a very good approximation for the
phase diagram of the AAT on the square lattice, since all the exact known
results concerning the location of the critical frontiers were recovered.
Unfortunately, as any Migdal-Kadanoff-like renormalization-group scheme, the
present one fails to detect some special features of the universality classes, in particular
the well-known result that the $F-P$ boundary for the IAT is a line of varying
critical exponent. In the following section we present our conclusions.

\section{Conclusions}

We have used a real-space renormalization group approach to obtain the
global phase diagram for the anisotropic ferromagnetic Ashkin-Teler model. Our findings
can be seen either as an {\it approximation} for the AAT on the square lattice, or
as an {\it exact} result for the hierarquical lattice generated by the scheme presented in
Fig. 1 \cite{berk,kauf}. In the former case, the resulting phase diagrams
reproduces qualitatively well the structure of those obtained previously
by other methods \cite{wulin,domany,bek}. Since many exact known results
for the square lattice are reproduced, we expect that the phase boundaries
(critical surfaces) represent a fairly good approximation for the exact ones.
It also important to note that this simple approach gives non-trivial
critical exponents, although it presents some discrepancies with respect to
the universality class. In particular, the line of varying critical exponent
is not reproduced. We believe that the results for the universality class
can be improved if one considers sistematically larger clusters
($b \rightarrow \infty$). However, such an enterprise is beyond the
purpose of the present work, which was motivated by a relatively simple
method to obtain reliable phase diagrams.

\vskip 2\baselineskip
{\large\bf Acknowledgments}

\vskip \baselineskip
\noindent
We are grateful to CNPq and Pronex/MCT (Brazilian research funding foundations)
for partial financial support.

\newpage
\centerline{{\large\bf Figure Captions}}

\vskip 2\baselineskip
\noindent
{\bf Fig. 1:} Self-dual graph used for the generation of the hierarchical lattice
considered in this work. The corresponding scaling factor is $b=2$.

\vskip 2\baselineskip
\noindent
{\bf Fig. 2:} Phase diagram in $(t_{1}^D,t_{2}^D,t_{3}^D)$ space.

\vskip 2\baselineskip
\noindent
{\bf Fig. 3:} Phase diagram in $(X,Y,Z)$ space. The $F_\tau$
boundaries are obtained by a reflection of the $F_\sigma$
boundaries on the $Z=0$ plane.

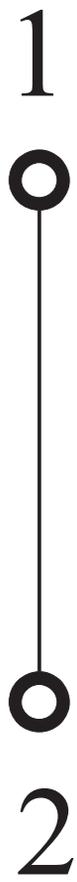
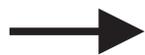
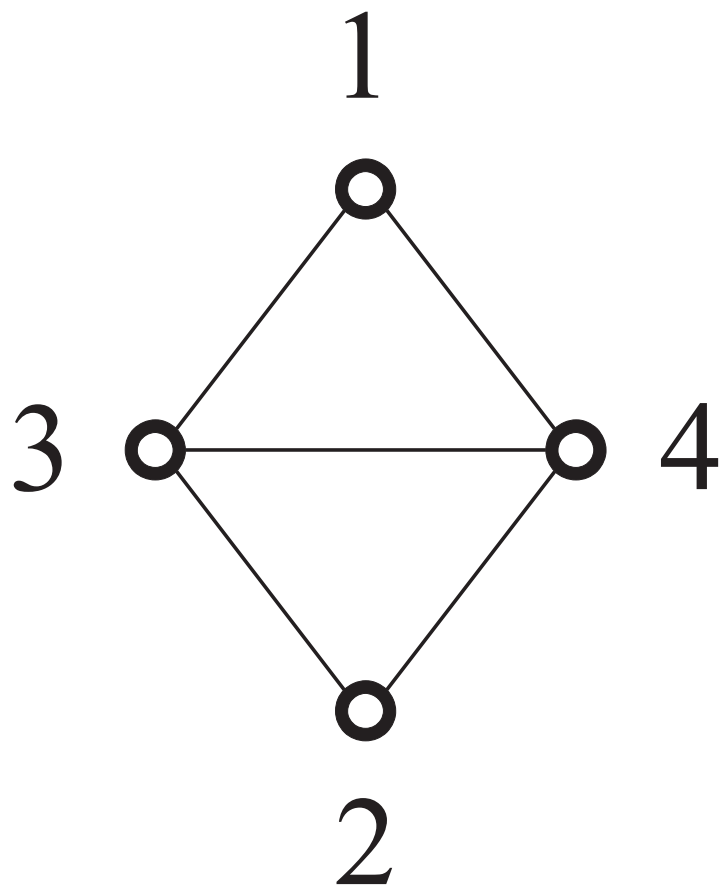
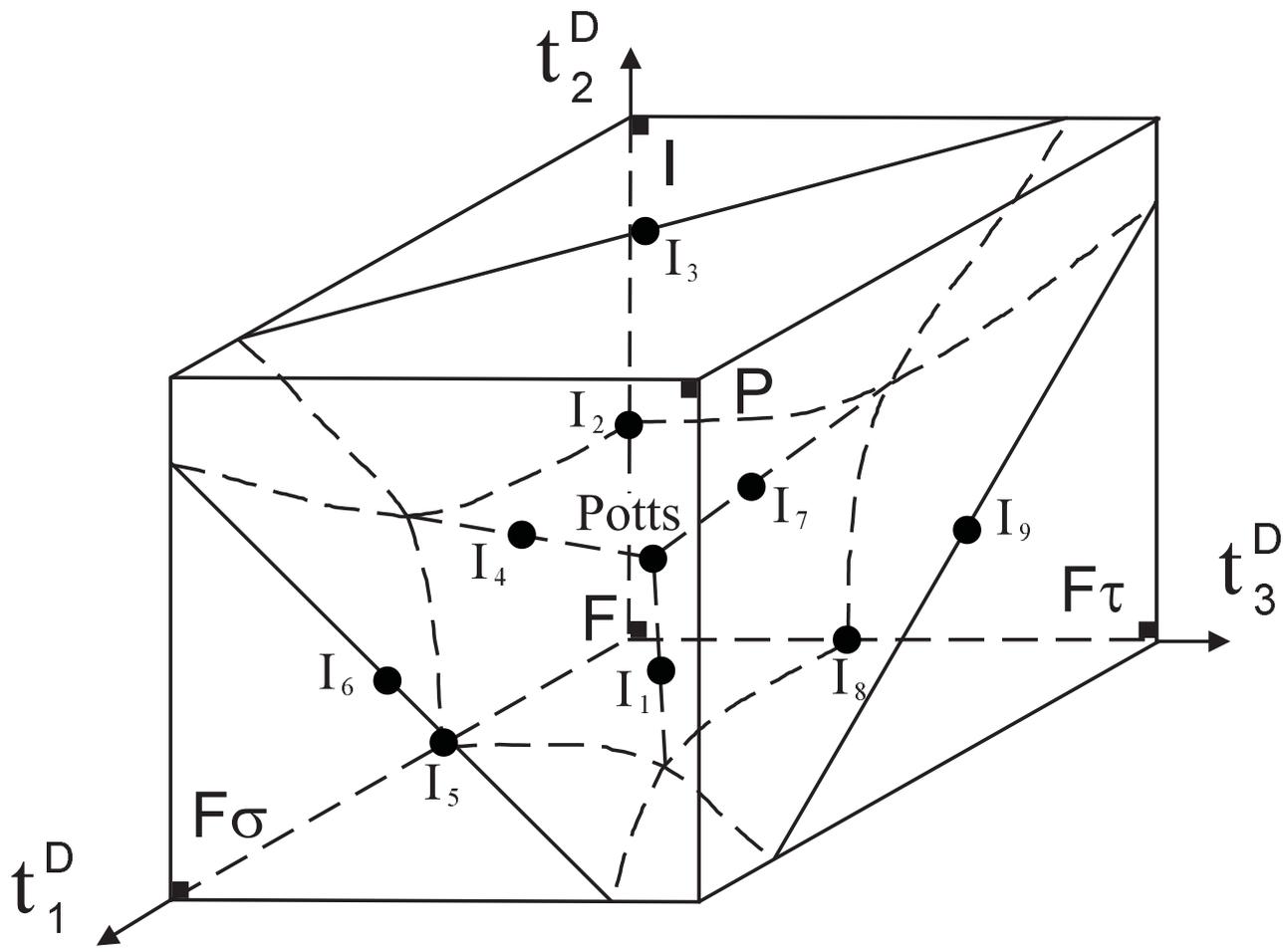
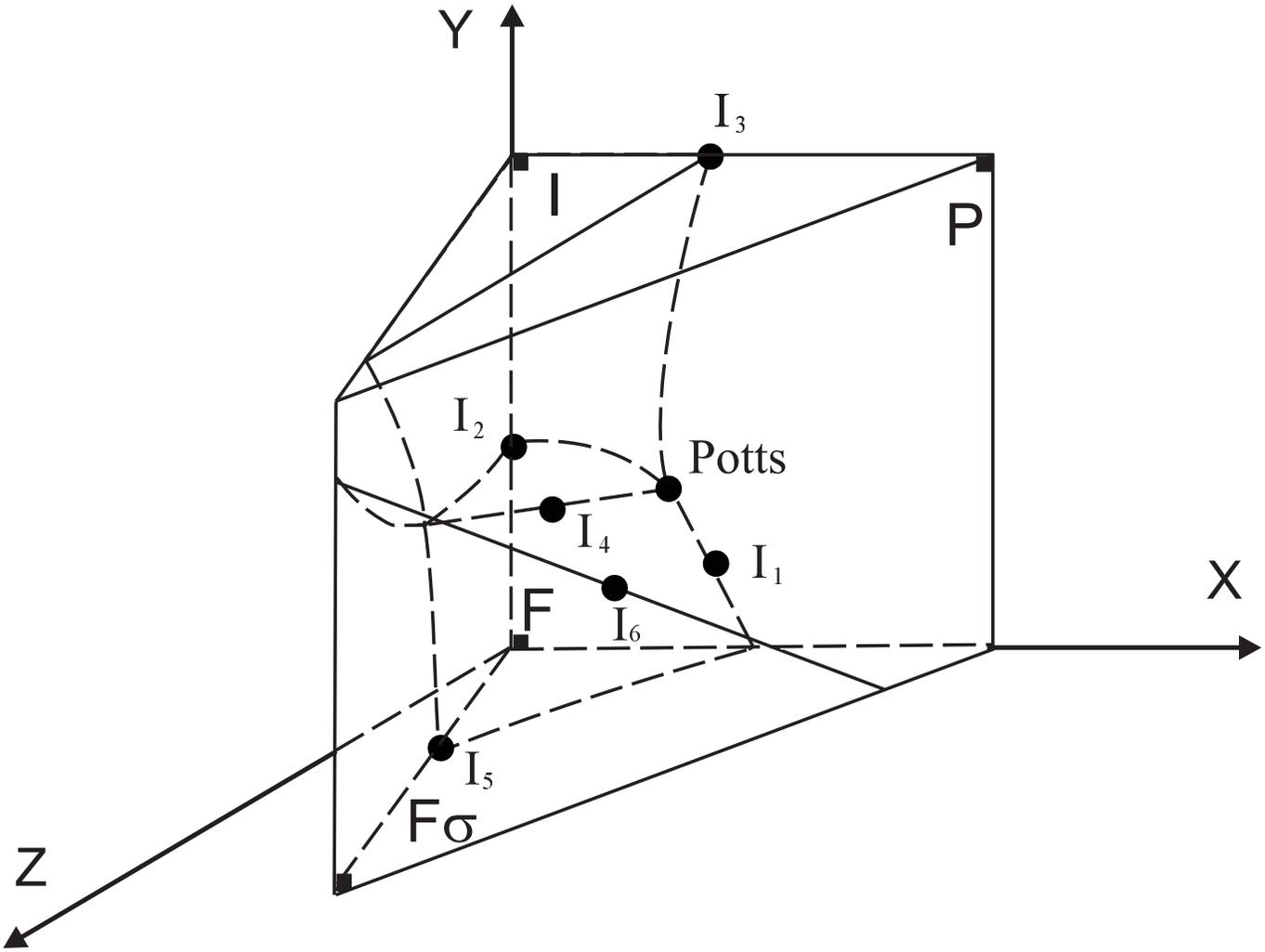
\end{document}